\documentclass[11pt]{article}
\usepackage{amssymb}
\usepackage{amsmath}
\usepackage{amsfonts}

\begin{document}

\newtheorem{theorem}{Theorem}
\newtheorem{df}[theorem]{Definition}
\newtheorem{lemma}[theorem]{Lemma}
\newtheorem{cor}[theorem]{Corollary}
\newtheorem{prop}[theorem]{Proposition}
\newtheorem{ex}{Example}
\newtheorem{conjecture}{Conjecture}
\newtheorem{open}[conjecture]{Open Problem}
\newtheorem{fact}[conjecture]{Fact}

\newcommand{\qed}{$\square$}

\newenvironment{proofofth}[1]{%
  \noindent{\it Proof of Theorem #1.\ }}{%
  \hspace*{\fill}\qed
  \vspace{2ex}}

\newenvironment{proofoflem}[1]{%
  \noindent{\it Proof of Lemma #1.\ }}{%
  \hspace*{\fill}\qed
  \vspace{2ex}}

\newenvironment{proof}{%
  \noindent{\it Proof.\ }}{%
  \hspace*{\fill}\qed
  \vspace{2ex}}

\newenvironment{sketchofproof}{%
  \noindent{\it Proof (Sketch).\ }}{%
  \hspace*{\fill}\qed
  \vspace{2ex}}

\hyphenation{boo-le-an}

\renewcommand{\thefootnote}{\fnsymbol{footnote}}

\newcommand{\mPhi}{{\mit\Phi}}
\newcommand{\mPsi}{{\mit\Psi}}

\newcommand{\bmpi}{\mbox{\boldmath $\pi$}}
\newcommand{\bmOmega}{\mbox{\boldmath $\Omega$}}
\newcommand{\bmPhi}{\mbox{\boldmath $\Phi$}}
\newcommand{\bmphi}{\mbox{\boldmath $\phi$}}
\newcommand{\bmPsi}{\mbox{\boldmath $\Psi$}}
\newcommand{\bmpsi}{\mbox{\boldmath $\psi$}}
\newcommand{\bmtheta}{\mbox{\boldmath $\theta$}}
\newcommand{\bmchi}{\mbox{\boldmath $\chi$}}
\newcommand{\bmA}{\mbox{\boldmath $A$}}
\newcommand{\bmd}{\mbox{\boldmath $d$}}
\newcommand{\bme}{\mbox{\boldmath $e$}}
\newcommand{\bmF}{\mbox{\boldmath $F$}}
\newcommand{\bmf}{\mbox{\boldmath $f$}}
\newcommand{\bmI}{\mbox{\boldmath $I$}}
\newcommand{\bmk}{\mbox{\boldmath $k$}}
\newcommand{\bmM}{\mbox{\boldmath $M$}}
\newcommand{\scbmM}{\mbox{\scriptsize\boldmath $M$}}
\newcommand{\bmp}{\mbox{\boldmath $p$}}
\newcommand{\bmq}{\mbox{\boldmath $q$}}
\newcommand{\bmR}{\mbox{\boldmath $R$}}
\newcommand{\bmr}{\mbox{\boldmath $r$}}
\newcommand{\bmS}{\mbox{\boldmath $S$}}
\newcommand{\bms}{\mbox{\boldmath $s$}}
\newcommand{\bmv}{\mbox{\boldmath $v$}}
\newcommand{\bmW}{\mbox{\boldmath $W$}}
\newcommand{\bmw}{\mbox{\boldmath $w$}}
\newcommand{\bmX}{\mbox{\boldmath $X$}}
\newcommand{\bmx}{\mbox{\boldmath $x$}}
\newcommand{\bmY}{\mbox{\boldmath $Y$}}
\newcommand{\bmy}{\mbox{\boldmath $y$}}
\newcommand{\bmZ}{\mbox{\boldmath $Z$}}
\newcommand{\bmz}{\mbox{\boldmath $z$}}

\newcommand{\NP}{\mbox{NP}}
\newcommand{\coNP}{\mbox{co-NP}}
\newcommand{\BPP}{\mbox{BPP}}
\newcommand{\BQP}{\mbox{BQP}}
\newcommand{\PP}{\mbox{PP}}
\newcommand{\PSPACE}{\mbox{PSPACE}}
\newcommand{\EXP}{\mbox{EXP}}
\newcommand{\NEXP}{\mbox{NEXP}}
\newcommand{\PnumberP}{\mbox{P}^{\mbox{\scriptsize \#P}}}
\newcommand{\IP}{\mbox{IP}}
\newcommand{\MIP}{\mbox{MIP}}
\newcommand{\QIP}{\mbox{QIP}}
\newcommand{\QMIP}{\mbox{QMIP}}
\newcommand{\MA}{\mbox{MA}}
\newcommand{\AM}{\mbox{AM}}
\newcommand{\QMA}{\mbox{QMA}}
\newcommand{\QAM}{\mbox{QAM}}
\newcommand{\PCP}{\mbox{PCP}}
\newcommand{\QOC}{\mbox{QOC}}
\newcommand{\CPOC}{\mbox{CPOC}}
\newcommand{\QIOC}{\mbox{QIOC}}
\newcommand{\NQP}{\mbox{NQP}}
\newcommand{\coCP}{\mbox{co-C}_{=}\mbox{P}}
\newcommand{\PrQSPACE}{\mbox{PrQSPACE}}

\newcommand{\MAP}{{\rm MAP}}

\newcommand{\tr}{{\rm tr}}
\newcommand{\rank}{{\rm rank}}

\newcommand{\init}{\psi_{\rm init}}
\newcommand{\final}{\psi_{\rm final}}
\newcommand{\bra}[1]{\langle #1 |}
\newcommand{\ket}[1]{| #1 \rangle}
\newcommand{\sbra}[1]{\langle #1 |}
\newcommand{\sket}[1]{| #1 \rangle}
\newcommand{\ketbra}[1]{| #1 \rangle \langle #1 |}
\newcommand{\braket}[2]{\langle #1 | #2 \rangle}
\newcommand{\abs}[1]{\vert #1 \vert}
\newcommand{\norm}[1]{\Vert #1 \Vert}
\newcommand{\snorm}[1]{\Vert #1 \Vert}
\newcommand{\trnorm}[1]{\Vert #1 \Vert_{\rm tr}}
\newcommand{\dnorm}[1]{\Vert #1 \Vert_{\diamond}}

\newcommand{\ent}{{\rm ent}}

\title{\Large\bf Quantum Certificate Verification:\\ Single versus Multiple Quantum Certificates}

\author{
{\large \hspace*{-1ex} $\mbox{\bf Hirotada Kobayashi}^{\ast \dagger}$ \hspace*{-1ex}}\\
{\tt \hspace*{-2ex} hirotada@is.s.u-tokyo.ac.jp \hspace*{-2ex}} 
\and
{\large \hspace*{-1ex} $\mbox{\bf Keiji Matsumoto}^{\dagger}$ \hspace*{-1ex}}\\
{\tt \hspace*{-2ex} keiji@qci.jst.go.jp \hspace*{-2ex}}
\and
{\large \hspace*{-1ex} $\mbox{\bf Tomoyuki Yamakami}^{\ddagger}$ \hspace*{-1ex}}\\
{\tt \hspace*{-2ex} yamakami@site.uottawa.ca \hspace*{-2ex}}
}

\date{2 October 2001}

\maketitle

\footnotetext[1]{
\noindent
Department of Information Science,
        Graduate School of Science, The University of Tokyo,
        7-3-1 Hongo, Bunkyo-ku, Tokyo 113-0033, Japan.}
\footnotetext[2]{
\noindent
Quantum Computation and Information Project, ERATO,
Japan Science and Technology Corporation,
        5-28-3 Hongo, Bunkyo-ku, Tokyo 113-0033, Japan.}
\footnotetext[3]{
\noindent
School of Information Technology and Engineering, 
University of Ottawa,
150 Louis Pasteur, P.O. Box 450, Stn A,
Ottawa, Ontario, Canada K1N 6N5.\\
This work was done while he was visiting the ERATO QCI project offices
between May and August, 2001.
} 

\begin{abstract}
The class $\MA$
consists of languages that can be efficiently verified
by classical probabilistic verifiers using a single classical certificate,
and the class $\QMA$
consists of languages that can be efficiently verified
by quantum verifiers using a single quantum certificate.
Suppose that a verifier receives not only one
but multiple certificates.
In the classical setting, it is obvious that
a classical verifier with multiple classical certificates
is essentially the same with the one with a single classical certificate.
However, in the quantum setting where a quantum verifier
is given a set of quantum certificates in tensor product form
(i.e. each quantum certificate is not entangled with others), 
the situation is different, because the quantum verifier
might utilize the structure of the tensor product form.
This suggests a possibility of another hierarchy of complexity classes,
namely the $\QMA$ hierarchy.
From this point of view, we extend the definition of $\QMA$
to $\QMA(k)$ for the case quantum verifiers use $k$ quantum certificates,
and analyze the properties of $\QMA(k)$.

To compare the power of $\QMA(2)$ with that of $\QMA(1) = \QMA$,
we show one interesting property of ``quantum indistinguishability''.
This gives a strong evidence that $\QMA(2)$ is more powerful than $\QMA(1)$.
Furthermore, we show that, for any fixed positive integer $k \geq 2$,
if a language $L$ has a one-sided bounded error $\QMA(k)$ protocol
with a quantum verifier using $k$ quantum certificates,
$L$ necessarily has a one-sided bounded error $\QMA(2)$ protocol
with a quantum verifier using only two quantum certificates.
\end{abstract}

\section{Introduction}

The class $\MA$ \cite{Bab85, BabMor88, Bab92}
is a randomized generalization of the class $\NP$.
A quantum analogue of $\MA$
was apparently first discussed by Knill \cite{Kni96},
was later studied by Kitaev \cite{Kit99},
and was named $\QMA$ by Watrous \cite{Wat00}.
Intuitively,
$\MA$ is the class of languages that can be efficiently verified
by classical probabilistic verifiers using a single classical certificate,
and $\QMA$ is the one that can be efficiently verified
by quantum verifiers using a single quantum certificate.

Consider a situation that a verifier receives not only one
but multiple certificates.
In the classical setting, it is obvious that
using multiple classical certificates does not increase
the power of the verifier and
the classical verifier with multiple classical certificates
remains the same in the computational power
with the one with a single classical certificate.
However, in the quantum setting where a quantum verifier
is given a set of quantum certificates unentangled each other,
the situation is different, because the quantum verifier
might utilize the fact that each certificate is not entangled with others.
From this point of view, we extend the definition of $\QMA$
to $\QMA(k)$
for the case quantum verifiers use $k$ quantum certificates,
and analyze the properties of $\QMA(k)$.

The most natural and important question to ask is
how much stronger the quantum verifier becomes
using two quantum certificates instead of one,
or in other words,
how much amount of help it is for the verifier
to know the existence of a tensor product structure.
For this question, this paper gives a strong evidence
of the quantum verifier with two quantum certificates
being much stronger than the one with only one quantum certificate.
More precisely, we show somewhat a surprising result stated as follows.
\begin{theorem}
Suppose one of the following two is true
for given $2n$ qubits certificate $\ket{\mPsi}$:
\begin{itemize}
 \item [(a)]  $\ketbra{\mPsi}\in {\sf H}_0$,
$
{\sf H}_0=\{\ketbra{\mPsi} \mid
        \exists\ket{\psi},\ket{\phi} : n \mbox{ qubits pure states},
        \; \ket{\mPsi} = \ket{\psi}\otimes\ket{\phi}\},
$
 \item [(b)]  $\ketbra{\mPsi}\in {\sf H}^{\varepsilon}_{1}$,
$
{\sf H}^{\varepsilon}_1=\{\ketbra{\mPsi} \mid
        \forall\ket{\psi},\ket{\phi} : n \mbox{ qubits pure states},
        \; F(\ketbra{\mPsi}, \ketbra{\psi}\otimes\ketbra{\phi})\leq 1-\varepsilon\}.
$
\end{itemize}
Then, for any $0 \leq \varepsilon \leq 1 - 2^{-n/2}$,
there is no quantum measurement better than
the trivial strategy where one guesses at random without 
any operation at all.
\label{indistinguishability}
\end{theorem}
Note that $F(\cdot, \cdot)$ in the statement above represents
the fidelity, and the formal definition of the fidelity is given
in Section \ref{Quantum Fundamentals}.
This theorem holds true even if we replace ${\sf H}^{\varepsilon}_{1}$
by the set of maximally entangled states
(see Lemma \ref{no-distinguish} in Section \ref{Two Certificates vs. One Certificate}). 
We emphasize that this theorem is a quantum information theoretical one,
and
the indistinguishability stated in this theorem
holds as far as we obey the law of quantum physics.
Thus this theorem suggests something like
``{\it quantum indistinguishability}'' so to speak.

Although it is given above a strong evidence
that using two quantum certificates is much different from
using only one quantum certificate,
this paper also points out that
the situation might change in comparing $\QMA(3)$ with $\QMA(2)$.
Let us say that
a language $L$ has a one-sided bounded error $\QMA(k)$ protocol
if there exists a quantum polynomial-time verifier $V$
and a polynomially bounded function $p \geq 1$
such that, for every input $x$ of length $n$,
(i) if $x \in L$, there exists a set of $k$ quantum certificates
which causes $V$ accept $x$ with probability $1$,
and (ii) if $x \not\in L$,
then for any set of $k$ quantum certificates,
$V$ accepts $x$ with probability at most $1 - 1/p(n)$.
In fact, we prove the following property.
\begin{theorem}
Let $L$ be a language having a one-sided bounded error $\QMA(3)$ protocol.
Then $L$ has a one-sided bounded error $\QMA(2)$ protocol.
\label{3 to 2}
\end{theorem}
A key idea to prove this theorem is to make use of the fact
that there is no entanglement
between two certificates given in $\QMA(2)$ protocols.
Let $\ket{C_{1}}$, $\ket{C_{2}}$, and $\ket{C_{3}}$ be three certificates
given in a $\QMA(3)$ protocol.
If two certificates $\ket{D_{1}}$ and $\ket{D_{2}}$
given in the corresponding $\QMA(2)$ protocol are of the form
$\ket{D_{1}} = \ket{C_{1}} \otimes \ket{C_{3}}$
and
$\ket{D_{2}} = \ket{C_{2}} \otimes \ket{C_{3}}$,
it is obvious that we can simulate the $\QMA(3)$ protocol
by the $\QMA(2)$ protocol.
We use the Controlled-Swap operator
and construct an efficient test for the decomposability
of $\ket{D_{1}}$ and $\ket{D_{2}}$ into the form above.
We also show that, in the one-sided error cases,
this Controlled-Swap test is optimal in view of error probability
to check this decomposability.

Actually, Theorem \ref{3 to 2} can be generalized to the following theorem:
\begin{theorem}
For any positive integer $k$ and any $r \in \{0, 1, 2\}$, 
let $L$ be a language having a one-sided bounded error $\QMA(3k+r)$ protocol.
Then $L$ has a one-sided bounded error $\QMA(2k+r)$ protocol.
\label{3k to 2k}
\end{theorem}
Applying this theorem repeatedly, we obtain the following theorem:
\begin{theorem}
For any fixed positive integer $k$,
let $L$ be a language having a one-sided bounded error $\QMA(k)$ protocol.
Then $L$ has a one-sided bounded error $\QMA(2)$ protocol.
\end{theorem}
One important point to be mentioned concerning this theorem is that
we do not know if $L$ has an arbitrary small one-sided bounded error
$\QMA(2)$ protocol,
even if $L$ has an exponentially small one-sided error
$\QMA(k)$ protocol.
The situations are similar even in the cases of Theorem \ref{3 to 2}
and Theorem \ref{3k to 2k}.
It remains open for $k \geq 2$
whether running polynomially many copies of the the $\QMA(k)$ protocol
in parallel (i.e. parallel repetition)
reduces the error probability to be exponentially small.
For this reason,
there still remains the possibility,
even in the one-sided bounded error cases,
that each $\QMA(k)$ does not collapse to the other
and forms the $\QMA$ hierarchy.

The remainder of this paper is organized as follows.
In Section \ref{Quantum Fundamentals} we briefly review
basic notations, definitions, and properties
in quantum computation and quantum information theory,
which are used in this paper.
In Section \ref{Definitions} we give a formal definition of our model.
In Section \ref{Two Certificates vs. One Certificate}
we show an interesting property of ``quantum indistinguishability''.
This gives a strong evidence
that using two quantum certificates
is more powerful than using only one quantum certificate.
In Section \ref{k Certificates vs. Two Certificates}
we show that, for any fixed positive integer $k \geq 2$,
if a language $L$ has a one-sided bounded error $\QMA(k)$ protocol
with a quantum verifier using $k$ quantum certificates,
$L$ necessarily has a one-sided bounded error $\QMA(2)$ protocol
with a quantum verifier using only two quantum certificates.
Finally 
we conclude with Section \ref{Conclusion and Open Problems},
which summarizes our results and mentions a number of open problems
related to our model.

\section{Quantum Fundamentals}
\label{Quantum Fundamentals}

Here we briefly review basic notations and definitions
in quantum computation and quantum information theory.
Detailed descriptions are, for instance, in \cite{Gru99, NieChu00}.

A {\it pure state} is described by a unit vector
in some Hilbert space. In particular,
an $n$-dimensional pure state is a unit vector $\ket{\psi}$ in ${\Bbb C}^{n}$.
Let $\{ \ket{e_{1}}, \ldots, \ket{e_{n}} \}$
be an orthonormal basis for ${\Bbb C}^{n}$.
Then any pure state in ${\Bbb C}^{n}$ can be described as
$\sum_{i=1}^{n} \alpha_{i} \ket{e_{i}}$
for some
$\alpha_{1}, \ldots, \alpha_{n} \in {\Bbb C},
\sum_{i=1}^{n} |\alpha_{i}|^{2} = 1$.

A {\it mixed state} is a classical probability distribution
$(p_{i}, \ket{\psi_{i}}), 0 \leq p_{i} \leq 1, \sum_{i}p_{i} = 1$
over pure states $\ket{\psi_{i}}$.
This can be interpreted as being in the pure state
$\ket{\psi_{i}}$ with probability $p_{i}$.
A mixed state is often described in the form of a density matrix
$\rho = \sum_{i}p_{i} \ketbra{\psi_{i}}$.
Any density matrix is positive semidefinite and has trace $1$.

If a unitary transformation $U$ is applied to a state $\ket{\psi}$,
the state becomes $U \ket{\psi}$,
or in the form of density matrices,
a state $\rho$ changes to $U \rho U^{\dagger}$ after $U$ is applied.

One of the important operations to density matrices is
the {\it trace-out} operation.
Given a density matrix $\rho$ over ${\cal H} \otimes {\cal K}$,
the state after tracing out ${\cal K}$ is a density matrix over ${\cal H}$
described by
\[
\tr_{\cal K} \rho
= \sum_{i=1}^{n} (I_{\cal H} \otimes \bra{e_{i}})
\rho (I_{\cal H} \otimes \ket{e_{i}})
\]
for any orthonormal basis $\{ \ket{e_{1}}, \ldots, \ket{e_{n}} \}$ of ${\cal K}$,
where $n$ is the dimension of ${\cal K}$
and $I_{\cal H}$ is the identity operator over ${\cal H}$.
To perform this operation on some part of a quantum system
gives a partial view of the quantum system with respect to the remaining part.

For any mixed state with its density matrix $\rho$ over ${\cal H}$,
there is a pure state $\ket{\psi}$ in ${\cal H} \otimes {\cal K}$
for the Hilbert space ${\cal K}$ of $\dim({\cal K}) = \dim({\cal H})$
such that $\ket{\psi}$ is a {\it purification} of $\rho$,
that is, $\tr_{\cal K} \ketbra{\psi} = \rho$.

One of the important concepts in quantum physics is a {\it measurement}.
Any collection of linear operators $\{ A_{1}, \ldots, A_{k} \}$
satisfying $\sum_{i=1}^{k} A_{i}^{\dagger} A_{i} = I$ defines a measurement.
If a system is in a pure state $\ket{\psi}$,
such a measurement results in $i$ with probability $\norm{A_{i} \ket{\psi}}^{2}$,
and the state becomes $A_{i} \ket{\psi} / \norm{A_{i} \ket{\psi}}$.
If a system is in a mixed state with its density matrix $\rho$,
the result $i$ is observed with probability $\tr(A_{i} \rho A_{i}^{\dagger})$,
and the state after the measurement is with its density matrix
$A_{i} \rho A_{i}^{\dagger} / \tr(A_{i} \rho A_{i}^{\dagger})$.
Let us write $M_{i} = A_i^{\dagger}A_i$ for each $i$.
Then the measurement $\{A_1,\ldots, A_k\}$ on $\rho$ results in $i$
with probability $\tr (M_i\rho)$.
Statistics of results of the measurement is decided by
$\bmM=\{M_1,\ldots ,M_k\}$,
and this set is called a {\it positive operator valued measure (POVM)}.
Formally, a POVM is defined to be a set of operators
$\bmM=\{M_1,\ldots ,M_k\}$ satisfying
(i) $M_i$ is a non-negative hermitian matrix
and (ii) $\sum_{i=1}^k M_i=I$.
For any POVM $\bmM$, there is a quantum mechanical measurement
such that the probability of the measurement results in $i$ is
equal to $\tr (M_i\rho)$. 
Therefore we may allow a little abuse of the term ``measurement''
instead of using the term ``POVM''. 
A special class of measurements are {\it projection} or {\it von Neumann} measurements
where $\{ A_{1}, \ldots, A_{k} \}$ is a collection of orthonormal projections.
In this scheme,
an observable is a decomposition of ${\cal H}$ into orthogonal subspaces
${\cal H}_{1}, \ldots, {\cal H}_{k}$, that is,
${\cal H} = {\cal H}_{1} \oplus \cdots \oplus {\cal H}_{k}$.
More mathematically rigorous descriptions of quantum measurements are,
for example, in \cite{Hol82, Oza84}.
It is important to note
that two mixed states having same density matrix
cannot be distinguished at all by any measurement. 


The {\it trace norm}
of the linear operator $A$ is defined by
\[
\norm{A}_{\tr} = \frac{1}{2} \tr \sqrt{A^{\dagger}A}.
\]
In general, the trace norm $\norm{\rho - \sigma}_{\tr}$
gives an appropriate measure of distance
between two density matrices $\rho$ and $\sigma$.

Another important measure between two density matrices $\rho$ and $\sigma$ 
is the {\it fidelity} $F(\rho, \sigma)$ defined by
\[
F(\rho, \sigma) = \tr \sqrt{\sqrt{\rho} \sigma \sqrt{\rho}}.
\]
For any density matrices $\rho, \sigma$,
$0 \leq F(\rho, \sigma) \leq 1$ is satisfied,
and $F(\rho, \sigma) = 1$ if and only if $\rho = \sigma$.

The following two are important properties on the trace norm
and the fidelity.

\begin{theorem}[\cite{AhaKitNis98}]
Let $\bmp^{\scbmM} = (p_{1}^{\scbmM}, \ldots, p_{m}^{\scbmM}),
\bmq^{\scbmM} = (q_{1}^{\scbmM}, \ldots, q_{m}^{\scbmM})$
be the probability distributions generated by a POVM $\bmM$
on mixed states with density matrices $\rho, \sigma$, respectively.
Then, for any POVM $\bmM$,
$1/2 \abs{\bmp^{\scbmM} - \bmq^{\scbmM}}
\leq
\norm{\rho - \sigma}_{\tr}$,
where
$\abs{\bmp^{\scbmM} - \bmq^{\scbmM}}
= \sum_{i=1}^{m} \abs{p_{i}^{\scbmM} - q_{i}^{\scbmM}}$.
\label{trnorm and probability}
\end{theorem}

\begin{theorem}[\cite{FucGra99}]
For any density matrices $\rho$ and $\sigma$,
\[
1 - F(\rho, \sigma)
\leq \norm{\rho - \sigma}_{\tr}
\leq \sqrt{1 - (F(\rho, \sigma))^{2}}.
\]
\label{trnorm and fidelity}
\end{theorem}

\vspace{-4mm}

\section{Definitions}
\label{Definitions}

\subsection{Polynomial-time Uniformly Generated Families of Quantum Circuits}

First we review the concept of polynomial-time uniformly generated families
of quantum circuits according to \cite{KitWat00}.

A family $\{Q_{x} \}$ of quantum circuits is said to be
{\it polynomial-time uniformly generated} if there exists a deterministic procedure
that, on each input $x$, outputs a description of $Q_{x}$
and runs in time polynomial in $n = |x|$.
For simplicity, we assume all input strings are over the alphabet $\Sigma = \{ 0, 1 \}$.
It is assumed that the circuits in such a family are composed only of gates
in the Shor basis \cite{Sho96}: Hadamard gates, $\sqrt{\sigma_{z}}$ gates, and Toffoli gates.
Furthermore, it is assumed that the number of gates in any circuit is not more than
the length of the description of that circuit,
therefore $Q_{x}$ must have size polynomial in $n$.
For convenience,
we may identify a circuit $Q_{x}$ with the unitary operator it induces.

As is mentioned in \cite{KitWat00, Wat00, KobMat01},
to permit non-unitary quantum circuits, in particular,
to permit measurements at any timing in the computation
does not change the computational power of the model.
See \cite{AhaKitNis98} for a detailed description
of the equivalence of the unitary and non-unitary quantum circuit models.

\subsection{Quantum Verifier with Multiple Quantum Certificates}

Watrous \cite{Wat00} defined the class $\QMA$ in terms of quantum circuits.
Here we extend this definition of $\QMA$
and define the class $\QMA(k)$
for the case quantum verifiers use $k$ quantum certificates.

Let $k$ be the number of certificates.
For each input $x \in \Sigma^{\ast}$, of length $n = |x|$,
each quantum certificate $\ket{C_{i}}$ is a quantum pure state
consists of $q_{{\cal M}_{i}}(n)$ qubits.
Without loss of generality we assume that
$q_{{\cal M}_{1}} = \cdots = q_{{\cal M}_{k}} = q_{\cal M}$
holds for some polynomially bounded function
$q_{\cal M} \colon {\Bbb Z}^{+} \rightarrow {\Bbb N}$.

Besides $k q_{\cal M} (n)$ qubits for the certificates,
we have $q_{\cal V} (n)$ qubits called {\it private qubits}
in our quantum circuit.
Hence, the whole system of our quantum circuit consists of
$q_{\cal V}(n) + k q_{\cal M}(n)$ qubits.
All the private qubits are initialized to the $\ket{0}$ state,
and one of the private qubits is designated as the output qubit.

A $(q_{\cal V}, q_{\cal M})$-restricted quantum verifier $V$
is a polynomial-time computable mapping of the form
$V \colon \Sigma^{\ast} \rightarrow \Sigma^{\ast}$,
where $\Sigma = \{ 0, 1 \}$ is the alphabet set.
For any $x$ of length $n$,
$V(x)$ is a description of a polynomial-time uniformly generated
quantum circuit acting on
$q_{\cal V}(n) + k q_{\cal M}(n)$ qubits.

The probability that $V$ accepts the input $x$ is defined
to be the probability that an observation of the output qubit
(in the $\{ \ket{0}, \ket{1} \}$ basis) yields $1$,
after the circuit $V(x)$ is applied
to the state
$\ket{0^{q_{\cal V}(n)}}
\otimes \ket{C_{1}} \otimes \cdots \otimes \ket{C_{k}}$.


\begin{df}
For a positive integer $k$
and functions $a,b \colon {\Bbb Z}^{+} \rightarrow [0,1]$,
a language $L$ is in $\QMA(k,a,b)$
if there exist polynomially bounded functions
$q_{\cal V}, q_{\cal M} \colon {\Bbb Z}^{+} \rightarrow {\Bbb N}$
and a $(q_{\cal V}, q_{\cal M})$-restricted quantum verifier $V$
such that, for any $x$ of length $n$,
\begin{itemize}
\item[(i)]
if $x \in L$,
there exists a set of quantum certificates
$\ket{C_{1}}, \ldots, \ket{C_{k}}$
of $q_{\cal M}(n)$ qubits
such that, given $\ket{C_{1}}, \ldots, \ket{C_{k}}$,
$V$ accepts $x$ with probability at least $a(n)$,
\item[(ii)]
if $x \not\in L$,
given any set of quantum certificates
$\ket{C'_{1}}, \ldots, \ket{C'_{k}}$
of $q_{\cal M}(n)$ qubits,
$V$ accepts $x$ with probability at most $b(n)$.
\end{itemize}
\label{QMAdef}
\end{df}

For convenience, we say that a language $L$ has a one-sided bounded error
$\QMA(k)$ protocol iff $L$ is in $\QMA(k, 1, 1-1/p)$
for some polynomially bounded function $p \geq 1$.
We also write $\QMA(k)$ in short if it is not confusing to omit
the arguments corresponding to the error probabilities.

Note that the definition of the class $\QMA(k, a, b)$
is closely related to quantum multi-prover interactive proof systems
which was introduced by Kobayashi and Matsumoto \cite{KobMat01},
the model without any prior entanglement among provers.
Of particular interest are
$1$-message quantum $k$-prover interactive proof systems,
which can be shown equivalent in view of computational power
to the model of quantum verifiers with $k$ quantum certificates.  
In fact, the class $\QMA(k, a, b)$ is equal to $\QMIP(k, 1, a, b)$,
the class of languages having $1$-message quantum $k$-prover
interactive proof systems with two-sided error probability $(a,b)$,
where the provers do not share any prior entanglement before the computation.

\section{Two Quantum Certificates versus One Quantum Certificate}
\label{Two Certificates vs. One Certificate}

First we focus on the relation between $\QMA(2)$ and $\QMA(1)$.
Obviously, $\QMA(1, a, b) \subseteq \QMA(2, a, b)$ is satisfied
for any functions $a, b$.
Concerning whether the other side of inclusion holds,
it is natural to consider
the simulation of the protocol with two quantum certificates by
using only one quantum certificate.
In this section, we show the strong implication that 
the simulation is impossible.

Suppose we have a quantum subroutine which
answers which of (a) and (b) is true for given certificate
$\ket{\mPsi} \in {\cal H}^{\otimes 2}$ of $2n$ qubits,
where ${\cal H}$ is the Hilbert space which consists of $n$ qubits:
\begin{itemize}
 \item [(a)]  $\ketbra{\mPsi} \in {\sf H}_{0}$,
$
{\sf H}_0=\{\ketbra{\mPsi} \mid
        \exists\ket{\psi},\ket{\phi} : n \mbox{ qubits pure states},
        \; \ket{\mPsi} = \ket{\psi}\otimes\ket{\phi}\},
$
 \item [(b)]  $\ketbra{\mPsi} \in {\sf H}^{\varepsilon}_1$,
$
{\sf H}^{\varepsilon}_1=\{\ketbra{\mPsi} \mid
        \forall\ket{\psi},\ket{\phi} : n \mbox{ qubits pure states},
        \; F(\ketbra{\mPsi}, \ketbra{\psi}\otimes\ketbra{\phi})
\leq 1-\varepsilon \}.
$
\end{itemize}
As for the certificate $\ket{\mPsi}$ which does not satisfy (a) nor (b),
this subroutine may answer (a) or (b) at random.
If the subroutine answers that $\ket{\mPsi}$ satisfies (b),
the verifier of the $\QMA(1)$ protocol rejects.
Otherwise, using the certificate $\ket{\mPsi}$,
the quantum verifier fulfills the same verification procedure
as the one in the original $\QMA(2)$ protocol.
It seems to the authors that there is no other way to simulate
two quantum certificates by only one quantum certificate.
Therefore, the authors conjecture that QMA(1) is strictly smaller than
QMA(2), because 
this kind of subroutines cannot be realized by any physical method,
which can be proven as follows.

In fact, we prove stronger lemma
which claims that states in tensor product form cannot be distinguished 
even from maximally entangled states by any physical operation.
Here, the state $\rho=\ketbra{\mPsi}$ is said
to be {\it maximally entangled} 
if $\ket{\mPsi}$ can be written by 
\begin{eqnarray*}
 \ket{\mPsi}=\sum_{i=1}^k {\alpha_i} \ket{e_i}\otimes\ket{f_i},
    \; |\alpha_i|^2=\frac{1}{d},\; i=1,\ldots, d,
\end{eqnarray*}
where 
$d = 2^n$ is the dimension of ${\cal H}$
and $\{\ket{e_1},\ldots,\ket{e_d}\}$, $\{\ket{f_1},\ldots,\ket{f_d}\}$
are orthonormal bases of ${\cal H}$ \cite{BenBerPopSch96}.
Among all states,
maximally entangled states are farthest away from states
in tensor product form, and  
$\min \{F(\ketbra{\mPsi}, \ketbra{\phi}\otimes\ketbra{\psi})\}
=1/\sqrt{d} = 2^{-n/2}$.

\begin{lemma}
\label{no-distinguish}
 Suppose one of the following two is true for the certificate
$\ket{\mPsi} \in {\cal H}^{\otimes 2}$ of $2n$ qubits:
\begin{itemize}
 \item [(a)]  $\ketbra{\mPsi}\in {\sf H}_0$,
$
{\sf H}_0 =\{\ketbra{\mPsi} \mid
        \exists\ket{\psi},\ket{\phi} \in {\cal H} : n \mbox{\rm { qubits pure states}},
        \; \ket{\mPsi} = \ket{\psi}\otimes\ket{\phi}\}.
$
 \item [(b)]  $\ketbra{\mPsi}\in {\sf H}_1$,
    ${\sf H}_1 =\{ \ketbra{\mPsi} \mid \ket{\mPsi}
                \mbox{\rm { is maximally entangled}} \}$.
\end{itemize}
Then, no quantum measurement is better 
than the trivial strategy where one guesses at random without 
any operation at all.
\end{lemma}
\begin{proof}
Let $\bmM = \{M_0, M_1\}$ be a POVM on given $\ketbra{\mPsi}$.
With $\bmM$ we conclude $\ketbra{\mPsi}\in{\sf H}_i$
if $\bmM$ results in $i$ ($i=0,1$).
Let ${\rm P}^{\scbmM}_{i\rightarrow j}(\ketbra{\mPsi})$
denote the probability that
$\ketbra{\mPsi}\in{\sf H}_j$ is concluded while 
$\ketbra{\mPsi}\in{\sf H}_i$ is true.
We want to find the measurement which  minimizes
${\rm P}^{\scbmM}_{0\rightarrow 1}(\ketbra{\mPsi})$
keeping the other side of error small enough.
More precisely, we want to evaluate ${\cal E}$ defined and bounded as follows.
\begin{eqnarray*}
 {\cal E}
& \stackrel{\rm def}{=} &
\min_{\scbmM}\left\{\max_{\rho\in {\sf H}_0}
 {\rm P}^{\scbmM}_{0\rightarrow 1}(\rho)
\;\left|\;
\max_{\rho\in{\sf H}_1}
{\rm P}^{\scbmM}_{1\rightarrow 0}(\rho)\leq\delta \right.\right\}\\
& \geq &
\min_{\scbmM}\left\{\int_{\rho\in {\sf H}_0}
 {\rm P}^{\scbmM}_{0\rightarrow 1}(\rho)\mu_0({\rm d}\rho)
 \;\left|\;
\int_{\rho\in{\sf H}_1}
{\rm P}^{\scbmM}_{1\rightarrow 0}(\rho)\mu_1({\rm d}\rho)\leq\delta
\right.\right\}
\\
&=&\min_{\scbmM}\left\{
 {\rm P}^{\scbmM}_{0\rightarrow 1}\left(\int_{\rho\in {\sf H}_0}\rho\mu_0({\rm d}\rho)\right)
 \;\left|\;
{\rm P}^{\scbmM}_{1\rightarrow 0}\left(\int_{\rho\in{\sf H}_1}\rho\mu_1({\rm d}\rho)\right)
                    \leq\delta \right.\right\},
\end{eqnarray*} 
where each $\mu_i$ is an arbitrary probability measure in ${\sf H}_i$.
This means that ${\cal E}$ is larger than
the error probability of distinguishment of 
two states  $\int_{\rho\in {\sf H}_0}\rho\mu_0({\rm d}\rho)$
and $\int_{\rho\in {\sf H}_1}\rho\mu_1({\rm d}\rho)$.
Furthermore, there exists $\mu_i$ such that 
\begin{eqnarray}
 \int_{\rho\in {\sf H}_0}\rho\mu_0({\rm d}\rho)
=\int_{\rho\in {\sf H}_1}\rho\mu_1({\rm d}\rho)
=\frac{1}{d^2} I_{{\cal H}^{\otimes 2}}.
\label{I div d-square}
\end{eqnarray}
Here, $\mu_0$ is a uniform distribution on the set
$\{\ketbra{e_i}\otimes\ketbra{e_j}\}_{i=1}^d{}_{j=1}^d$,
that is, $\mu_{0}(\{ \ketbra{e_{i}} \otimes \ketbra{e_{j}} \}) = 1/d^{2}$
for each $i, j$,
where $\{\ket{e_1},\ldots\ket{e_d}\}$ is an orthonormal basis of ${\cal H}$.
Similarly, $\mu_1$ is a uniform distribution on the set
$\{\ket{g_{n,m}}\bra{g_{n,m}}\}_{n=1}^d{}_{m=1}^d$,
that is, $\mu_{1}(\{ \ketbra{g_{n,m}} \}) = 1/d^{2}$
for each $n,m$,
where 
\begin{eqnarray*}
 \ket{g_{n,m}}=\frac{1}d
\sum_j e^{2\pi\sqrt{-1} jn/d}
\ket{e_j}\otimes\ket{e_{(j+m) \bmod d}}.
\end{eqnarray*} 
This $\{\ket{g_{1,1}}, \ldots, \ket{g_{d,d}}\}$
is an orthonormal basis of ${\cal H}^{\otimes 2}$
\cite{BenBraCreJozPreWoo93}.
Thus we have the assertion from (\ref{I div d-square}).
\end{proof}

From Lemma \ref{no-distinguish}, we can easily show
Theorem \ref{indistinguishability}.

\section{$\bmk$ Quantum Certificates versus Two Quantum Certificates}
\label{k Certificates vs. Two Certificates}

In the last section we gave a strong evidence
that using two quantum certificates is much different from
using only one quantum certificate.
Here we point out that
the situation might change if we compare the case of
using $k \geq 2$ quantum certificates with the case of
using only two quantum certificates.
The C-SWAP algorithm described below is the key idea
of our claim.

\subsection{Utilization of Controlled-Swap}

Given a pair of $n$ qubits mixed states $\rho, \sigma$
of the form $\rho \otimes \sigma$,
consider the following algorithm,
which we call {\it C-SWAP algorithm}.
A restricted version of this algorithm
with an input to be a pair of $n$ qubits pure states
is utilized in \cite{BuhCleWatWol01}.

We prepare quantum registers ${\bf B}$, ${\bf R}_1$, and ${\bf R}_2$.
${\bf B}$ consists of only one qubit,
both of ${\bf R}_1$ and ${\bf R}_2$ consist of $n$ qubits,
and all the qubits in ${\bf B}$, ${\bf R}_1$, and ${\bf R}_2$
are initially set to the $\ket{0}$ state.

\begin{itemize}
\item[]
{\bf C-SWAP Algorithm}
\begin{enumerate}
\item
Set $\rho, \sigma$ in ${\bf R}_1, {\bf R}_2$, respectively.
\item
Apply the Hadamard transformation $H$ to ${\bf B}$.
\item
Apply controlled-swap operator on ${\bf R}_1$ and ${\bf R}_2$
with using ${\bf B}$ as a control qubit.
That is, 
swap the contents of ${\bf R}_1$ and ${\bf R}_2$
if ${\bf B}$ contains $1$,
and do nothing if ${\bf B}$ contains $0$. 
\item
Apply the Hadamard transformation $H$ to ${\bf B}$
and accept if ${\bf B}$ contains $0$.
\end{enumerate}
\end{itemize}

\begin{prop}
The probability that the input $\rho, \sigma$ is accepted
in the C-SWAP algorithm is exactly $1/2 + \tr (\rho \sigma)/2$.
\end{prop}

\begin{proof}
Let ${\cal B}, {\cal R}_1, {\cal R}_2$ denote the Hilbert spaces
corresponding to the qubits in ${\bf B}, {\bf R}_1, {\bf R}_2$,
respectively.
Let $\rho = \sum_{i} p_{i} \ketbra{\phi_{i}}$
and $\sigma = \sum_{j} q_{j} \ketbra{\psi_{j}}$
be decomposition of $\rho$ and $\sigma$
with respect to the orthonormal bases $\{\ket{\phi_{i}}\}, \{\ket{\psi_{j}}\}$
of ${\bf R}_1, {\bf R}_2$, respectively.

We introduce the Hilbert spaces
${\cal S}_1 = l_2(\Sigma^n)$ and ${\cal S}_2 = l_2(\Sigma^n)$.
Then there exist purifications
$\ket{\phi} \in {\cal R}_1 \otimes {\cal S}_1$
and $\ket{\psi} \in {\cal R}_2 \otimes {\cal S}_2$
of $\rho$ and $\sigma$, respectively, such that
\[
\ket{\phi} = \sum_{i}\sqrt{p_i}\ket{\phi_{i}}\ket{\phi_{i}},\quad
\ket{\psi} = \sum_{j}\sqrt{q_j}\ket{\psi_{j}}\ket{\psi_{j}}.
\]

Now consider the following pure state
$\ket{\xi} \in {\cal B}
\otimes {\cal R}_1 \otimes {\cal S}_1
\otimes {\cal R}_2 \otimes {\cal S}_2$,
\[
\ket{\xi}
= \ket{0}\ket{\phi}\ket{\psi}
= \sum_{i,j}\sqrt{p_iq_j}
\ket{0}\ket{\phi_{i}}\ket{\phi_{i}}\ket{\psi_{j}}\ket{\psi_{j}}.
\]
The probability that the input pair of $\rho, \sigma$ is accepted
in the C-SWAP algorithm is
exactly equal to the probability of acceptance
when the C-SWAP algorithm is applied to $\ket{\xi}$
over the Hilbert space ${\cal B} \otimes {\cal R}_1 \otimes {\cal R}_2$.

If the C-SWAP algorithm is applied to $\ket{\xi}$,
we can easily see that the state
$\ket{\eta} \in {\cal B}
\otimes {\cal R}_1 \otimes {\cal S}_1
\otimes {\cal R}_2 \otimes {\cal S}_2$
before the final measurement of the output qubit is given by
\begin{eqnarray*}
\ket{\eta}
& = &
\frac{1}{2} \ket{0}
  \otimes \left(
            \sum_{i,j} \sqrt{p_i q_j}
              \left(
                \ket{\phi_{i}}\ket{\phi_{i}}\ket{\psi_{j}}\ket{\psi_{j}}
                + \ket{\psi_{j}}\ket{\phi_{i}}\ket{\phi_{i}}\ket{\psi_{j}}
              \right)
          \right)\\
& &
\hspace*{5ex}
+ \frac{1}{2} \ket{1}
  \otimes \left(
            \sum_{i,j} \sqrt{p_i q_j}
              \left(
                \ket{\phi_{i}}\ket{\phi_{i}}\ket{\psi_{j}}\ket{\psi_{j}}
                - \ket{\psi_{j}}\ket{\phi_{i}}\ket{\phi_{i}}\ket{\psi_{j}}
              \right)
          \right).
\end{eqnarray*}
Thus the probability of acceptance is $(1+t)/2$, where $t$ is given by
\begin{eqnarray*}
\lefteqn{
t =
\left(
  \sum_{i,j} \sqrt{p_i q_j}
    \ket{\phi_{i}}\ket{\phi_{i}}\ket{\psi_{j}}\ket{\psi_{j}},
  \sum_{i,j} \sqrt{p_i q_j}
    \ket{\psi_{j}}\ket{\phi_{i}}\ket{\phi_{i}}\ket{\psi_{j}}
  \right)
=
\sum_{i,j} p_i q_j
  \left(
    \ket{\phi_{i}}\ket{\psi_{j}},
    \ket{\psi_{j}}\ket{\phi_{i}}
  \right)
}\\
& &
=
\sum_{i,j} p_i q_j
  \braket{\phi_{i}}{\psi_{j}}\braket{\psi_{j}}{\phi_{i}}
=
\sum_{i} p_i \bra{\phi_{i}} \sigma \ket{\phi_{i}}
=
\sum_{i} p_i \tr (\sigma \ketbra{\phi_{i}})
=
\tr (\rho \sigma),
\end{eqnarray*}
where $(\cdot, \cdot)$ represents the inner product.
This completes the proof.
\end{proof}

\subsection{Reducing the Number of Quantum Certificates}

Now we consider reducing the number of quantum certificates $k$,
given a $\QMA(k)$ protocol of using $k$ quantum certificates.
First we consider simulating one-sided bounded error $\QMA(3)$ protocols
by one-sided bounded error $\QMA(2)$ protocols.

\begin{lemma}
For any polynomially bounded function
$p \colon {\Bbb Z}^{+} \rightarrow {\Bbb R}^{+}, p \geq 1$,
\[
\QMA(3, 1, 1 - 1/p) \subseteq \QMA \bigl(2, 1, 1 - 1/(10p^2) \bigr).
\]
\label{Lem3to2}
\end{lemma}

\vspace{-5ex}
\begin{proof}
Let $L$ be a language in $\QMA(3, 1, 1 - 1/p)$.
Given a $\QMA(3, 1, 1 - 1/p)$ protocol for $L$,
we construct a $\QMA \bigl(2, 1, 1 - 1/(10p^2) \bigr)$ protocol for $L$
in the following way.

Let $V$ be the quantum verifier of the original $\QMA(3, 1, 1 - 1/p)$ protocol.
For the input $x$ of length $n$,
suppose that each of quantum certificates $V$ receives consists of $q_{\cal M}(n)$ qubits
and the number of private qubit of $V$ is $q_{\cal V}(n)$.
Let $U$ be the unitary transformation
which the original quantum verifier $V$ applies.
Our new quantum verifier $W$ of the $\QMA \bigl(2, 1, 1 - 1/(10p^2) \bigr)$ protocol
prepares quantum registers ${\bf R}_{1}, {\bf R}_{2}, {\bf S}_{1}, {\bf S}_{2}$
for quantum certificates
and quantum registers ${\bf V}, {\bf B}$ for private computation.
Each of ${\bf R}_{i}$ and ${\bf S}_{i}$ consists of $q_{\cal M}(n)$ qubits,
${\bf V}$ consists of $q_{\cal V}(n)$ qubits,
and ${\bf B}$ consists of a single qubit.
$W$ receives two quantum certificates $\ket{D_1}, \ket{D_2}$
of length $2q_{\cal M}(n)$,
which are expected to be of the form
\begin{equation}
\ket{D_1} = \ket{C_1} \otimes \ket{C_3},\quad
\ket{D_2} = \ket{C_2} \otimes \ket{C_3},
\label{DCC}
\end{equation}
where each $\ket{C_i}$ is the $i$th quantum certificate
which the original quantum verifier $V$ receives.
Of course, each $\ket{D_i}$ may not be of the form above
and the first and the second $q_{\cal M}(n)$ qubits of 
$\ket{D_i}$ may be entangled.
Let ${\cal V}$, ${\cal B}$, each ${\cal R}_{i}$, and each ${\cal S}_{i}$
be the Hilbert spaces corresponding to the quantum registers
${\bf V}, {\bf B}, {\bf R}_{i}, {\bf S}_{i}$, respectively.
$W$ runs the following protocol:

\vspace{5mm}

\begin{itemize}
\item[]
Verifier $W$'s Protocol
\begin{enumerate}
\item
Set the contents of the first $q_{\cal M}(n)$ qubits of $\ket{D_1}$ in ${\bf R}_{1}$,
and the contents of the second $q_{\cal M}(n)$ qubits of $\ket{D_1}$ in ${\bf S}_{1}$.\\
Set the contents of the first $q_{\cal M}(n)$ qubits of $\ket{D_2}$ in ${\bf R}_{2}$,
and the contents of the second $q_{\cal M}(n)$ qubits of $\ket{D_2}$ in ${\bf S}_{2}$.
\item
Do one of the following two tests at random.
\begin{itemize}
\item[2.1]
{\sc Separability test}:\\
Apply the C-SWAP algorithm over ${\cal B} \otimes {\cal S}_{1} \otimes {\cal S}_{2}$,
using quantum registers ${\bf B}, {\bf S}_{1}, {\bf S}_{2}$.\\
Accept if ${\bf B}$ contains 0, otherwise reject.
\item[2.2]
{\sc Consistency test}:\\
Apply $U$ over ${\cal V} \otimes {\cal R}_{1} \otimes {\cal R}_{2} \otimes {\cal S}_{1}$,
using quantum registers ${\bf V}, {\bf R}_{1}, {\bf R}_{2}, {\bf S}_{1}$.\\
Accept iff the result corresponds to
the acceptance computation of the original quantum verifier.
\end{itemize}
\end{enumerate}
\end{itemize}

\vspace{5mm}

\begin{itemize}
\item[(i)]
In case the input $x$ of length $n$ is in $L$:\\
In the original $\QMA(3, 1, 1 - 1/p)$ protocol for $L$,
there exist quantum certificates $\ket{C_1}, \ket{C_2}, \ket{C_3}$
which cause the original quantum verifier $V$ accept $x$ with probability $1$.
In the constructed $\QMA(2)$ protocol,
let the quantum certificates $\ket{D_1}, \ket{D_2}$ be of the form
$\ket{D_1} = \ket{C_1} \otimes \ket{C_3},
\ket{D_2} = \ket{C_2} \otimes \ket{C_3}$.
Then it is obvious that the constructed quantum verifier $W$
accepts $x$ with probability $1$.
\item[(ii)]
In case the input $x$ of length $n$ is not in $L$:\\
Consider any pair of quantum certificates $\ket{D'_1}, \ket{D'_2}$,
which are set in the pairs of the quantum registers
$({\bf R}_{1}, {\bf S}_{1}), ({\bf R}_{2}, {\bf S}_{2})$, respectively.
Let $\rho = \tr_{{\cal R}_{1}} \ketbra{D'_1}$
and $\sigma = \tr_{{\cal R}_{2}} \ketbra{D'_2}$.
Let $\varepsilon = 1 - 1/p(n)$ and
$\delta = (-1+2\varepsilon+4\sqrt{1+\varepsilon-\varepsilon^{2}})/5$.
The reason why we set $\delta$ at this value will be clear later in the item b.
\begin{itemize}
\item[a.]
In case $\tr (\rho \sigma) \leq \delta$:\\
In this case the probability $\alpha$
that the input $x$ is accepted in the {\sc Separability test}
is at most
\[
\alpha \leq \frac{1}{2} + \frac{\delta}{2}
=
\frac{2+\varepsilon+2\sqrt{1+\varepsilon-\varepsilon^{2}}}{5}
\leq
\frac{4+2\varepsilon-\varepsilon^{2}}{5}
=
1 - \frac{(1-\varepsilon)^{2}}{5},
\]
where the second inequality is from the fact $a+b \geq 2\sqrt{ab}, a \geq 0, b \geq 0$.
Thus the verifier $W$ accepts the input $x$ with probability
at most
\[
\frac{1}{2} + \frac{\alpha}{2}
\leq 1 - \frac{(1-\varepsilon)^{2}}{10}
= 1 - \frac{1}{10(p(n))^{2}}.
\]
\item[b.]
In case $\tr (\rho \sigma) > \delta$:\\
$\tr (\rho \sigma) > \delta$
means the maximum eigenvalue $\lambda$ of $\rho$ satisfies
$\lambda > \delta$.
Thus there exist pure states
$\ket{C'_1} \in {\cal R}_{1}$
and $\ket{C'_3} \in {\cal S}_{1}$
such that
\[
F(\ketbra{C'_1} \otimes \ketbra{C'_3}, \ketbra{D'_1})
> \sqrt{\delta},
\]
since $\rho = \tr_{{\cal R}_{1}} \ketbra{D'_1}$.
Similarly, the maximum eigenvalue of $\sigma$
is more than $\delta$
and there exist pure states
$\ket{C'_2} \in {\cal R}_{2}$
and $\ket{C'_4} \in {\cal S}_{2}$
such that
\[
F(\ketbra{C'_2} \otimes \ketbra{C'_4}, \ketbra{D'_2})
> \sqrt{\delta}.
\]
Thus, letting
$\ket{\phi} = \ket{C'_1} \otimes \ket{C'_3} \otimes \ket{C'_2} \otimes \ket{C'_4}$
and $\ket{\psi} = \ket{D'_1} \otimes \ket{D'_2}$,
we have
\[
F(\ketbra{\phi}, \ketbra{\psi})
> \delta.
\]
Therefore, from Theorem \ref{trnorm and fidelity} we have
\[
\trnorm{\ketbra{\phi} - \ketbra{\psi}}
\leq
\sqrt{1 - (F(\ketbra{\phi}, \ketbra{\psi}))^{2}}
<
\sqrt{1 - \delta^{2}}.
\]
With Theorem \ref{trnorm and probability}, this implies that,
the probability $\beta$ that the input $x$ is accepted in the {\sc Consistency test} is bounded by
\[
\beta < \varepsilon + \sqrt{1 - \delta^{2}},
\]
since given any quantum certificates $\ket{C'_1}, \ket{C'_2}, \ket{C'_3}$
the original quantum verifier $V$ accepts the input $x$
with probability at most $\varepsilon = 1 - 1/p(n)$.
Noticing that $\delta$ satisfies
\[
\frac{1}{2} + \frac{\delta}{2}
= \varepsilon + \sqrt{1 - \delta^{2}},
\]
we can see that
\[
\beta < 1 - \frac{(1-\varepsilon)^{2}}{5}.
\]
Thus the verifier $W$ accepts the input $x$ with probability
at most
\[
\frac{1}{2} + \frac{\beta}{2}
< 1 - \frac{(1-\varepsilon)^{2}}{10}
= 1 - \frac{1}{10(p(n))^{2}}.
\]
\end{itemize}
\end{itemize}
\end{proof}

\newpage

The following theorem is an immediate consequence of Lemma \ref{Lem3to2}.
\newcounter{theoremtmp}
\setcounter{theoremtmp}{\value{theorem}}
\setcounter{theorem}{1}
\begin{theorem}
Let $L$ be a language having a one-sided bounded error $\QMA(3)$ protocol.
Then $L$ has a one-sided bounded error $\QMA(2)$ protocol.
\end{theorem}
\setcounter{theorem}{\value{theoremtmp}}

Actually, Lemma \ref{Lem3to2} can be
generalized to the following lemma:
\begin{lemma}
For any fixed positive integer $k$, any $r \in \{0, 1, 2\}$,
and any polynomially bounded function
$p \colon {\Bbb Z}^{+} \rightarrow {\Bbb R}^{+}, p \geq 1$,
\[\QMA(3k+r, 1, 1 - 1/p)
\subseteq \QMA(2k+r, 1, 1 - 1/(10p^2)).
\]
\label{Lem3kto2k}
\end{lemma}

\begin{proof}
Let $L$ be a language in $\QMA(3k+r, 1, 1 - 1/p)$.
Given a $\QMA(3k+r, 1, 1 - 1/p)$ protocol for $L$,
we construct a $\QMA \bigl(2k+r, 1, 1 - 1/(10p^2) \bigr)$ protocol for $L$
in the following way.

Let $V$ be the quantum verifier
of the original $\QMA(3k+r, 1, 1 - 1/p)$ protocol.
For the input $x$ of length $n$,
suppose that each of quantum certificates $V$ receives
consists of $q_{\cal M}(n)$ qubits
and the number of private qubit of $V$ is $q_{\cal V}(n)$.
Let $U$ be the unitary transformation
which the original quantum verifer $V$ applies.
Our new quantum verifier $W$
of the $\QMA \bigl(2k+r, 1, 1 - 1/(10p^2) \bigr)$ protocol
prepares quantum registers
${\bf R}_{1,1}, \ldots, {\bf R}_{1,k},
{\bf R}_{2,1}, \ldots, {\bf R}_{2,k},
{\bf S}_{1,1}, \ldots, {\bf S}_{1,k},
{\bf S}_{2,1}, \ldots, {\bf S}_{2,k},
{\bf R}_{3,1}, \ldots, {\bf R}_{3,r},
{\bf S}_{3,1}, \ldots, {\bf S}_{3,r}$
for quantum certificates
and quantum registers ${\bf V}, {\bf B}$ for private computation.
Each of ${\bf R}_{i,j}$ and ${\bf S}_{i,j}$ consist of $q_{\cal M}(n)$ qubits,
${\bf V}$ consists of $q_{\cal V}(n)$ qubits,
and ${\bf B}$ consists of a single qubit.
$W$ receives $2k+r$ quantum certificates
$\ket{D_{1,1}}, \ldots, \ket{D_{1,k}},
\ket{D_{2,1}}, \ldots, \ket{D_{2,k}},
\ket{D_{3,1}}, \ldots, \ket{D_{3,r}}$
of length $2q_{\cal M}(n)$,
which are expected to be of the form
\[
\begin{array}{ccl}
\ket{D_{1,j_1}} & = & \ket{C_{j_1}} \otimes \ket{C_{2k+j_1}},\\
\ket{D_{2,j_1}} & = & \ket{C_{k+j_1}} \otimes \ket{C_{2k+j_1}},\\
\ket{D_{3,j_2}} & = & \ket{C_{3k+j_2}} \otimes \ket{0^{q_{\cal M}(n)}},\\
\end{array}
\]
for each $1 \leq j_1 \leq k, 1 \leq j_2 \leq r$,
where each $\ket{C_i}$ is the $i$th quantum certificate
which the original quantum verifier $V$ receives.
Let ${\cal V}$, ${\cal B}$, each ${\cal R}_{i,j}$, and each ${\cal S}_{i.j}$
be the Hilbert spaces corresponding to quantum registers
${\bf V}, {\bf B}, {\bf R}_{i,j}, {\bf S}_{i,j}$, respectively.
$W$ runs the following protocol:

\vspace{5mm}

\begin{itemize}
\item[]
Verifier $W$'s Protocol
\begin{enumerate}
\item
For each $i,j$,
set the contents of the first $q_{\cal M}(n)$ qubits of $\ket{D_{i,j}}$
in ${\bf R}_{i,j}$,
and the contents of the second $q_{\cal M}(n)$ qubits of $\ket{D_{i,j}}$
in ${\bf S}_{i,j}$.
\item
For each $j$, if ${\bf S}_{3,j}$ contains $1$ in some qubit, reject.
\item
Do one of the following two tests at random.
\begin{itemize}
\item[3.1]
{\sc Separability test}:\\
Apply the C-SWAP algorithm over
${\cal B} \otimes
({\cal S}_{1,1} \otimes \cdots \otimes {\cal S}_{1,k})
\otimes ({\cal S}_{2,1} \otimes \cdots \otimes {\cal S}_{2,k})$.
using the quantum register ${\bf B}$,
the $k$-tuple of quantum registers $({\bf S}_{1,1}, \ldots, {\bf S}_{1,k})$,
and the $k$-tuple of quantum registers $({\bf S}_{2,1}, \ldots, {\bf S}_{2,k})$.\\
Accpet if ${\bf B}$ contains 0, otherwise reject.
\item[3.2]
{\sc Consistency test}:\\
Apply $U$ over
${\cal V} \otimes
{\cal R}_{1,1} \otimes \cdots \otimes {\cal R}_{1,k}
\otimes {\cal R}_{2,1} \otimes \cdots \otimes {\cal R}_{2,k}
\otimes {\cal S}_{1,1} \otimes \cdots \otimes {\cal S}_{1,k}
\otimes {\cal R}_{3,1} \otimes \cdots \otimes {\cal R}_{3,r}
$,
using quantum registers
${\bf V},
{\bf R}_{1,1}, \ldots, {\bf R}_{1,k},
{\bf R}_{2,1}, \ldots, {\bf R}_{2,k},
{\bf S}_{1,1}, \ldots, {\bf S}_{1,k},
{\bf R}_{3,1}, \ldots, {\bf R}_{3,r}$.\\
Accept iff the result corresponds to
the acceptance computation of the original quantum verifier.
\end{itemize}
\end{enumerate}
\end{itemize}

\vspace{5mm}

With a similar argument to the proof of Lemma \ref{Lem3to2},
we can show that the above protocol is actually
a $\QMA \bigl(2k+r, 1, 1 - 1/(10p^2) \bigr)$ protocol for $L$.
\end{proof}

The following theorem is an immediate consequence of Lemma \ref{Lem3kto2k}.
\setcounter{theoremtmp}{\value{theorem}}
\setcounter{theorem}{2}
\begin{theorem}
For any positive integer $k$ and any $r \in \{0, 1, 2\}$, 
let $L$ be a language having a one-sided bounded error $\QMA(3k+r)$ protocol.
Then $L$ has a one-sided bounded error $\QMA(2k+r)$ protocol.
\end{theorem}
\setcounter{theorem}{\value{theoremtmp}}

Now we show that any one-sided bounded error $\QMA(k)$ protocol
can be simulated by a $\QMA(2)$ protocol
with one-sided bounded error.

\begin{lemma}
For any fixed positive integer $k$
and any polynomially bounded function
$p_1 \colon {\Bbb Z}^{+} \rightarrow {\Bbb R}^{+}, p_1 \geq 1$,
there exists a polynomially bounded function
$p_2 \colon {\Bbb Z}^{+} \rightarrow {\Bbb R}^{+}, p_2 \geq 1$
such that
\[\QMA(k, 1, 1 - 1/p_{1})
\subseteq \QMA(2, 1, 1 - 1/p_2).
\]
\end{lemma}

\begin{proof}
By applying Lemma \ref{Lem3kto2k} $c = O(\log_{3/2} k)$ times repeatedly,
we can easily obtain
\[
\QMA(k, 1, 1 - 1/p_1)
\subseteq \QMA(2, 1, 1 - 1/(10^{2^{c}-1}p_1^{2^{c}})),
\]
for some constant $c$.
Taking the polynomially bounded function
$p_2 = 10^{2^{c}-1}p_1^{2^{c}}$ completes the proof.
\end{proof} 

Thus we obtain the following theorem.
\setcounter{theoremtmp}{\value{theorem}}
\setcounter{theorem}{3}
\begin{theorem}
For any fixed positive integer $k$,
let $L$ be a language having a one-sided bounded error $\QMA(k)$ protocol.
Then $L$ has a one-sided bounded error $\QMA(2)$ protocol.
\end{theorem}
\setcounter{theorem}{\value{theoremtmp}}

\subsection{Optimality of C-SWAP Algorithm}

In the previous subsection, 
we showed how to simulate a one-sided bounded error $\QMA(3)$ protocol
by a one-sided bounded error $\QMA(2)$ protocol using the C-SWAP algorithm.
One might suspect that there is a better simulation than ours
to avoid the increase of the error probability in the simulation.
In this subsection, we show that the C-SWAP algorithm is optimal to check 
the decomposability of (\ref{DCC})
with one-sided error probability.

Let $\bmM=\{M_0, M_1\}$ be a POVM.
If the result of $\bmM$ is 1, 
the certificate $\ket{\mPsi}$ is judged
that $\ketbra{\mPsi}$ is in ${\sf H}_{0}$, where
\[
{\sf H}_0 =
\{\ketbra{\mPsi} \mid 
 \ket{\mPsi}=\ket{C_1}\ket{C_2}\ket{C_3}\ket{C_3},\;
  \ket{C_i}\in{\cal H}\},
\]
for the Hilbert space ${\cal H}$ of dimension $d$.
Our problem is to derive the optimal measurement 
for judging whether $\ketbra{\mPsi}\in{\sf H}_0$ or not.
Here we only consider one-sided error cases,
hence, the measurement must conclude
$\ketbra{\mPsi}\in{\sf H}_0$ with probability $1$ if
$\ketbra{\mPsi}\in{\sf H}_0$ is true.

Define $P_{\rm sym}$ to be a projection operator in ${\cal H}^{\otimes 2}$
whose image is
\begin{eqnarray*}
 {\rm span}\{\ket{e_i}, \ket{e_i}\ket{e_j}+\ket{e_j}\ket{e_i} 
 \;(1\leq i\leq d, 1\leq j\leq d)\}, 
\end{eqnarray*}
where $\{\ket{e_1},\ldots\ket{e_d}\}$ 
is an orthonormal basis of ${\cal H}$.

\begin{lemma}
In the one-sided error cases, 
the optimal measurement $\bmM=\{M_0, M_1\}$ to  
judge whether $\ketbra{\mPsi}\in {\sf H}_0$ or not is given by
\begin{eqnarray}
M_0= I_{{\cal H}^{\otimes 2}}\otimes P_{\rm sym},\quad M_1=1-M_0. 
\label{psym}
\end{eqnarray}
\end{lemma}
\begin{proof}
Since one-sided error is assumed,
if $\ketbra{\mPsi}\in {\sf H}_0$ is true,
the result of the measurement $\bmM$ must be always 1,
therefore, for all $\ketbra{\mPsi}\in{\sf H}_0$,
$
 \tr (M_0\ketbra{\mPsi})=1
$
is satisfied.
Thus,
for every $\ket{C_1}, \ket{C_2}, \ket{C_3} \in{\cal H}$,
\[
  M_0 \ket{C_1}\ket{C_2}\ket{C_3}\ket{C_3}
 =\ket{C_1}\ket{C_2}\ket{C_3}\ket{C_3}.
\]
Hence, we have 
\begin{eqnarray*}
  M_0 \geq  I_{{\cal H}^{\otimes 2}}\otimes P_{\rm sym},
\end{eqnarray*}
which implies that, for any density matrix $\rho$
over the Hilbert space ${\cal H}^{\otimes 2}$,
\[
 \tr (\rho M_0) \geq  \tr(\rho I_{{\cal H}^{\otimes 2}}\otimes P_{\rm sym}).
\]
This means that
the measurement (\ref{psym})
minimizes
the probability of concluding  $\ketbra{\mPsi}\in {\sf H}_0$ when
$\ketbra{\mPsi}\not\in {\sf H}_0$.
\end{proof}

It is easy to check that our C-SWAP algorithm realizes 
the optimal POVM (\ref{psym}), and we have the following theorem.
\begin{theorem}
In the one-sided error cases,
the C-SWAP algorithm is optimal in view of error probability
to check the decomposability of (\ref{DCC}).
\end{theorem}

\section{Conclusion and Open Problems}
\label{Conclusion and Open Problems}

This paper introduced the class $\QMA(k)$
in which the quantum verifier uses $k$ quantum certificates.
This suggests a possibility of another hierarchy of complexity classes,
namely the $\QMA$ hierarchy.
It was given a strong evidence that $\QMA(2)$ differs from $\QMA(1)$,
and was also shown that, for any fixed positive integer $k \geq 2$,
if a language $L$ has a one-sided bounded error $\QMA(k)$ protocol,
$L$ necessarily has a one-sided bounded error $\QMA(2)$ protocol.

A number of interesting problems remain open in this paper.

\begin{itemize}
\item
In the case of $\QMA(1)$ protocols,
we can easily see that a parallel repetition of the protocol works well
\cite{KitWat00, Wat00}.
For $k \geq 2$,
does a parallel repetition of polynomially many times
of the $\QMA(k)$ protocol
reduce the error probability to be exponentially small?
\item
Kitaev and Watrous \cite{KitWat00, Wat00, Wat01}
showed the $\PP$ upper bound for the class $\QMA(1)$.
Is $\QMA(k)$ also contained in $\PP$ for $k \geq 2$?
\item
Can a two-sided bounded error $\QMA(k)$ protocol
be modified to a one-sided bounded error one?
\item
Does $\QMA(k)$ collapse to the other for some $k$,
or do they form the $\QMA$ hierarchy?
\item
Suppose that the quantum certificates be prepared
by the provers isolated each other,
but the provers share prior entanglement (cf. \cite{Cle00}).
Then how the situation changes? 
\end{itemize}

\section*{Acknowledgement}

The authors would like to thank John Watrous
for providing the proof of $\QMA \subseteq \PP$.
The authors also thank Richard Cleve and Lance Fortnow
for their helpful comments on writing this paper.


\begin{thebibliography}{99}
\bibitem{AhaKitNis98}
D. Aharonov, A. Kitaev, and N. Nisan.
\newblock Quantum circuits with mixed states.
\newblock In {\it Proceedings of the 30th Annual ACM Symposium on Theory of Computing},
pages 20--30, 1998.
\bibitem{Bab85}
L. Babai.
\newblock Trading group theory for randomness.
\newblock In {\it Proceedings of the 17th Annual ACM Symposium on Theory of Computing},
pages 421--429, 1985.
\bibitem{Bab92}
L. Babai.
\newblock Bounded round interactive proofs in finite groups.
\newblock {\it SIAM Journal on Discrete Mathematics}, 5(1):88--111, 1992.
\bibitem{BabMor88}
L. Babai and S. Moran.
\newblock Arthur-Merlin games: a randomized proof system, and a hierarchy of complexity classes.
\newblock {\it Journal of Computer and System Sciences}, 36(2):254--276, 1988.
\bibitem{BenBerPopSch96}
C. H. Bennett, H. J. Bernstein, S. Popescu, and B. Schumacher.
\newblock Concentrating partial entanglement by local operations.
\newblock {\it Physical Review A}, 53(4):2046--2052, 1996.
\bibitem{BenBraCreJozPreWoo93}
C. H. Bennett, G. Brassard, C. Cr{\'e}peau, R. Jozsa, A. Peres, and W. K. Wootters.
\newblock Teleporting an unknown quantum state via dual classical and Einstein-Podolsky-Rosen channels.
\newblock {\it Physical Review Letters}, 70(13):1895--1899, 1993.
\bibitem{BuhCleWatWol01}
H. Buhrman, R. Cleve, J. Watrous, and R. de Wolf.
\newblock Quantum fingerprinting.
\newblock To appear in {\it Physical Review Letters}, 87(16), 2001.
\newblock Los Alamos Preprint Archive, quant-ph/0102001.
\bibitem{Cle00}
R. Cleve.
\newblock An entangled pair of provers can cheat.
\newblock Talk at the Workshop on Quantum Computation and Information, California Institute of Technology, November 2000.
\bibitem{FucGra99}
C. Fuchs and J. van de Graaf.
\newblock Cryptographic distinguishability measures for quantum-mechanical states.
\newblock {\it IEEE Transactions on Information Theory}, 45(4):1216--1227, 1999.
\bibitem{Gru99}
J. Gruska.
\newblock {\it Quantum Computing}.
\newblock McGraw-Hill, 1999.
\bibitem{Hol82}
A. S. Holevo.
\newblock {\it Probabilistic and Statistical Aspects of Quantum Theory}.
\newblock North-Holland, 1982.
\bibitem{Kit99}
A. Kitaev.
\newblock Quantum $\NP$.
\newblock Talk at the 2nd Workshop on Algorithms in Quantum Information Processing,
DePaul University, Chicago, January 1999.
\bibitem{KitWat00}
A. Kitaev and J. Watrous.
\newblock Parallelization, amplification, and exponential time simulation of quantum interactive proof systems.
\newblock In {\it Proceedings of the 32nd Annual ACM Symposium on Theory of Computing},
pages 608--617, 2000.
\bibitem{Kni96}
E. Knill.
\newblock Quantum randomness and nondeterminism.
\newblock Technical Report LAUR-96-2186, Los Alamos National Laboratory, 1996.
\bibitem{KobMat01}
H. Kobayashi and K. Matsumoto.
\newblock On the power of quantum multi-prover interactive proof systems.
\newblock Los Alamos Preprint Archive, cs.CC/0102013, 2001.
\bibitem{NieChu00}
M. A. Nielsen and I. L. Chuang.
\newblock {\it Quantum Computation and Quantum Information}.
\newblock Cambridge University Press, 2000.
\bibitem{Oza84}
M. Ozawa.
\newblock Quantum measuring processes of continuous observables.
\newblock {\it Journal of Mathematical Physics}, 25(1):79--87, 1984. 
\bibitem{Sho96}
P. W. Shor.
\newblock Fault-tolerant quantum computation.
\newblock In {\it Proceedings of the 37th Annual Symposium on Foundations of Computer Science},
pages 56--65, 1996.
\bibitem{Wat00}
J. Watrous.
\newblock Succinct quantum proofs for properties of finite groups.
\newblock In {\it Proceedings of the 41st Annual Symposium on Foundations of Computer Science},
pages 537--546, 2000.
\bibitem{Wat01}
J. Watrous.
\newblock Private communication, 2001.
\end{thebibliography}
\end{document}